\documentclass[aps,prb,twocolumn,amsmath,longbibliography]{revtex4-2}  
\usepackage{bbm}
\usepackage{mathrsfs}
\usepackage{amsmath}
\usepackage{amsfonts}
\usepackage{graphicx,epstopdf}
\usepackage{subfigure}
\usepackage{epsfig}
\usepackage{dcolumn}
\usepackage{natbib}
\usepackage{url}
\usepackage{bm}
\usepackage{color}
\usepackage{amssymb}
\usepackage{xcolor}
\usepackage{accents}
\usepackage{braket}
\usepackage[colorlinks=true,citecolor=blue,linkcolor=blue,anchorcolor=blue]{hyperref}

\begin{document}

\title{ Linearly Polarized Light-Induced Anomalous Hall Effect and Topological Phase Transitions in  Altermagnetic Topological Insulators}

\author{Yichen Liu$^{1}$, Tongshuai Zhu$^{2,\dagger}$, and Haijun Zhang$^{1,3,4,\dagger}$}
\email[]{zhanghj@nju.edu.cn ; tongshuaizhu@upc.edu.cn}

\affiliation{$^{1}$National Laboratory of Solid State Microstructures and School of Physics, Nanjing University, Nanjing 210093, China}
\affiliation{$^{2}$College of Science, China University of Petroleum (East China), Qingdao 266580, China}
\affiliation{$^{3}$Jiangsu Physical Science Research Center, Nanjing 210093, China}
\affiliation{$^{4}$Jiangsu Key Laboratory of Quantum Information Science and Technology, Nanjing University, China}

\date{\today}

\begin{abstract}
A recently identified class of collinear magnetic order, characterized by vanishing net magnetization yet unconventional spin splitting, known as altermagnets (AMs), has attracted significant research interest. Controlling the unconventional spin splitting and the associated band topology in AMs offers opportunities for realizing novel spin and topological transport phenomena. In this work, using Floquet engineering with periodically driven linearly polarized light (LPL), we explore light-induced control of an AM topological insulator. Remarkably, we find that AMs and conventional antiferromagnets (AFMs) exhibit distinct responses under LPL irradiation. Specifically, since LPL breaks neither time-reversal ($\mathcal{T}$) symmetry nor parity-time-reversal ($\mathcal{PT}$) symmetry, it is incapable of generating spin splitting or inducing an anomalous Hall effect (AHE) in conventional AFMs. In contrast, AMs intrinsically lack both $\mathcal{T}$ and $\mathcal{PT}$ symmetries. Their spin-up and spin-down bands are related by the combined symmetry of time reversal $\mathcal{T}$ and a crystal rotation. We show that LPL readily breaks these symmetries, thereby triggering a finite AHE exclusively in AMs. Furthermore, LPL can drive the AM topological insulator into a fully spin-polarized Chern insulating phase. Our findings not only provide a robust experimental scheme to distinguish AMs from conventional AFMs, but also establish a promising pathway toward dissipationless spintronic applications.

\end{abstract}

\maketitle

\section{Introduction}
Altermagnetism (AM) has recently emerged as an unconventional magnetic phase, distinct from conventional ferromagnets (FMs) and antiferromagnets (AFMs), attracting extensive attention\cite{smejkal2022emerging,smejkal2022beyond,smejkal2022giant,he2025evidence,attias2024intrinsic,reimers2024direct,ding2024large,zhou2025manipulation,mazin2023altermagnetism,krempasky2024altermagnetic,lee2024broken,osumi2024observation,leiviska2024anisotropy,reichlova2024observation,rial2024altermagnetic,ouassou2023dc,zhang2024finite,cheng2024orientation,sun2023andreev,fang2024quantum,qin2026layer,chen2025quasicrystalline,shao2025classification,li2025unconventional,lin2025coulomb,wang2025spinorbital,zhou2024crystal,ma2021multifunctional,wang2024electric,duan2025antiferroelectric,jiang2025metallic,xu2025altermagnetic,bai2024altermagnetism,song2025electrical,Liu2024Twisted,guo2025altermagnet,Gu2025Ferroelectric,wang2026hyperbolic,fu2025all,yamada2025metallic}. Conventional AFMs exhibit zero net magnetization and spin-degenerate  bands. In contrast, AMs are characterized by a time-reversal ($\mathcal{T}$) symmetry breaking while preserving its combination with specific crystalline symmetries. With negligible spin–orbit coupling (SOC), AMs can be naturally described by the spin group symmetry, which enforces vanishing macroscopic magnetization while permitting strongly anisotropic spin-split electronic band structures in momentum space\cite{hayami2019momentum,hayami2020bottom,smejkal2022giant,smejkal2022beyond,smejkal2022emerging,jiang2024enumeration,chen2024enumeration,xiao2024spin,Liu2022Spin,wang2026hyperbolic}.  By combining key features of both FMs and AFMs, AMs therefore provide a promising platform for spintronic applications. Beyond their magnetic properties, the intrinsic spin-split band structures of AMs also offer a platform for realizing nontrivial band topology\cite{li2025topological,wan2025altermagnetism,rao2024tunable,ma2024altermagnetic,antonenko2025mirror,qu2025altermagnetic,parshukov2025topological,fernandes2024topological,li2024creation,feng2025type,chen2025quantum,zhang2025quantized,gonzalez2025spin,gonzalez2025model}, which plays a central role in robust edge transport phenomena. As a result, the simultaneous manipulation of unconventional spin splitting and topological electronic states in AMs has recently attracted growing interest.

Among various approaches to controlling electronic structures, periodic light irradiation offers a powerful dynamical route known as Floquet engineering, enabling access to diverse light-induced phenomena\cite{lindner2011floquet,wang2013observation,yan2016tunable,huebener2017creating,zhou2016floquet,liu2018photoinduced,gomezleon2013floquet,rudner2013anomalous,ezawa2013photoinduced,grushin2014floquet,mahmood2016selective,oka2019floquet,mikami2016brillouin,eckardt2015high,ma2021floquet,wang2017line,bomantara2018simulation,nag2021anomalous,ghosh2022systematic,trevisan2022bicircular,bao2022light,rudner2020band,bielinski2025floquetbloch,li2024npl,merboldt2025floquetgraphene,ma2025metasurface,choi2025observation,fan2025floquet,li2024floquet,Zhu2024Manipulating,Zhan2022Floquet,oka2009photovoltaic,kitagawa2011transport,mciver2020light,xu2021light,kong2022floquet,ning2022photoinduced,wang2018light,yap2017computational,zhu2023Floquet,bukov2015universal,Chen2018Floquet,min2026transition}. Recently, light-induced manipulation of electronic and magnetic properties in AMs has attracted growing interest, such as light-induced odd-parity magnetism\cite{zhu2025floquet,huang2025light,liu2025light,Li2025Stacking}, light-induced higher-order SOC\cite{ghorashi2025dynamical}, Floquet engineering spin triplet states\cite{Fu2026Floquet,fuSciPostPhys.20.2.059}, circularly polarized light (CPL)-induced quantum anomalous Hall (QAH) Effect  \cite{Zou2025Floquet} and so on\cite{ganguli2025tunable,qin2026anomalous,wang2026emerging,Pan2025Floquet,libo2025floquet,yarmohammadi2025anisotropic,yarmohammadi2026xxx,xt23-9pnv}. In contrast to previous proposals where light-induced  anomalous Hall effects (AHE) predominantly rely on CPL to break  $\mathcal{T}$ symmetry and  parity-time-reversal ($\mathcal{PT}$) symmetry\cite{wang2013observation,oka2009photovoltaic,kitagawa2011transport,mciver2020light,xu2021light,kong2022floquet,ning2022photoinduced,wang2018light,yap2017computational,zhu2023Floquet}, linearly polarized light (LPL) is typically ineffective in nonmagnetic and conventional antiferromagnetic materials. AMs circumvent this limitation  owing to their intrinsically broken $\mathcal{T}$ and $\mathcal{PT}$ symmetries. Moreover, AMs themselves can host nontrivial band topology, and AM topological insulator phases have been explored \cite{wan2025altermagnetism,rao2024tunable,ma2024altermagnetic,antonenko2025mirror,qu2025altermagnetic,parshukov2025topological,fernandes2024topological,li2024creation,feng2025type,chen2025quantum,zhang2025quantized,gonzalez2025spin,gonzalez2025model}. Despite these advances, the interplay between LPL driving and the underlying topology and magnetism of AMs remains largely unexplored.

\begin{figure}
    \centering
    \includegraphics[width=\linewidth]{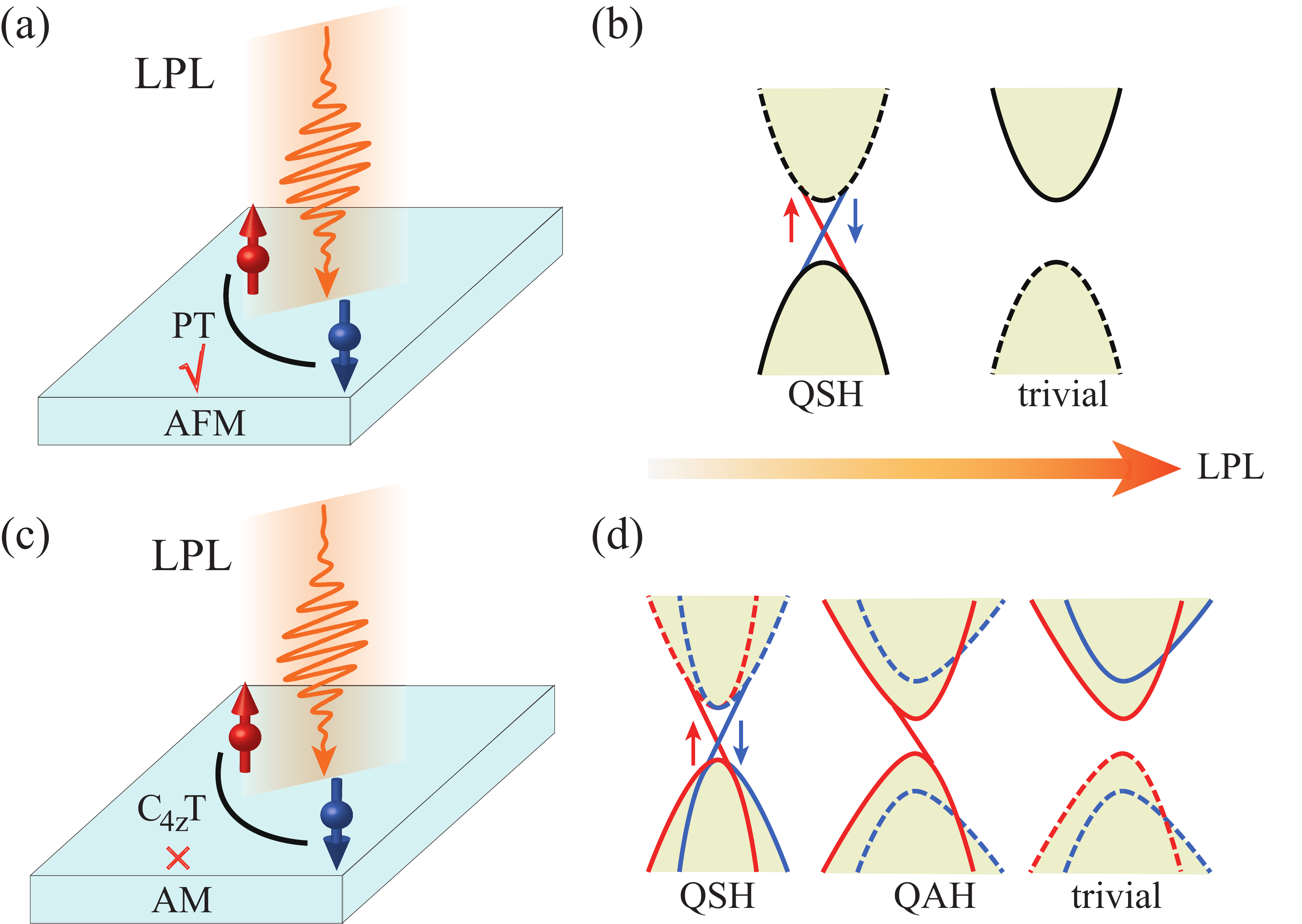}
    \caption{Schematic illustrations of linearly polarized light (LPL) control on the altermagnet (AM) and antiferromagnet (AFM). (a) In conventional AFM, the spin-up and spin-down sublattices are connected by $\mathcal{PT}$ symmetry, and incident LPL does not break $\mathcal{PT}$ symmetry. (b) As the light intensity increases, the two-dimensional AFM quantum spin Hall insulator transitions into a trivial insulator. The black lines indicate that the energy bands for spin-up and spin-down are degenerate, and the dashed lines represent band inversion. (c) In the AM, spin-up and spin-down sublattices are connected by $C_{4z}\mathcal{T}$ symmetry, and incident LPL breaks this symmetry. (d) As the light intensity increases, the two-dimensional AM quantum spin Hall insulator first transitions into a spin-polarized Chern insulator and then into a trivial insulator. The red and blue lines represent the energy bands for spin-up and spin-down, respectively.}
    \label{fig1}
\end{figure}

In this work, we employ a four-band model on a two-dimensional square lattice with $d$-wave spin splitting, and derive the effective Hamiltonian under LPL irradiation within Floquet theory. We calculate and analyze the band structures, anomalous Hall conductivity (AHC), and topological phase transitions induced by LPL in out-of-plane AFMs and $d$-wave AMs. We reveal that LPL cannot break the $\mathcal{PT}$ symmetry connecting the spin-up and spin-down sublattices [Fig.~\ref{fig1}(a)], and therefore cannot lift the spin degeneracy. As the intensity of the LPL increases, it  drives the AFM QSH insulator into a trivial insulator, as schematically illustrated in Fig.~\ref{fig1}(b).  In contrast,  we find that in the AM, where the spin-up and spin-down sublattices are connected by $C_{4z}\mathcal{T}$ symmetry, LPL can break this symmetry[Fig.~\ref{fig1}(c)]. Consequently, the spin-up and spin-down bands do not close and reopen simultaneously as the LPL intensity increases. This allows the AM QSH insulator to undergo a transition to a spin-polarized Chern insulator under LPL irradiation, as shown in Fig.~\ref{fig1}(d). Furthermore, we observe that the AHE becomes anisotropic with respect to the polarization direction of the LPL, enabling sign reversal upon polarization rotation. Our work not only provides an experimentally feasible means to distinguish between AMs and AFMs but also reveals the different topological phase transitions in AFM and AM topological insulators under LPL irradiation.

\section{Model and Method}
To investigate the manipulation of AM via Floquet engineering with LPL, we first employ an out-of-plane  $d$-wave AM tight-binding model and subsequently utilize Floquet theory to derive the effective model under LPL irradiation.

\subsection{The d-wave AM tight-binding model}
We employ a four-band two-dimensional out-of-plane $d$-wave  AM on a square lattice \cite{ma2024altermagnetic,gonzalez2025spin}, in which  the orbitals are  $p_{z\uparrow}$, $\frac{1}{\sqrt{2}} (d_{xz\uparrow} + i d_{yz\uparrow})$, $p_{z\downarrow}$, and $\frac{1}{\sqrt{2}} (d_{xz\downarrow} - i d_{yz\downarrow})$. The corresponding Hamiltonian is similar to the Bernevig-Hughes-Zhang  model in momentum space and is expressed as follows:
\begin{equation}
    H(\mathbf{k}) = \begin{pmatrix} 
        H_{\uparrow}(\mathbf{k}) & 0 \\ 
        0 & H_{\downarrow}(\mathbf{k}) 
    \end{pmatrix},
    \label{eq1}
\end{equation}
where $H_{\downarrow}(k_x,k_y)=H_{\uparrow}(k_y,k_x)$, $H_{\uparrow}(\mathbf{k})=\sum_i{d_i}^{\uparrow}(\mathbf{k})\tau_{i}$ with $i=x,y,z$, $\tau_{i}$ denote the Pauli matrices acting on the orbital subspace, $d_{x}^{\uparrow}(\mathbf{k})=-v  \sin a k_x$, $d_{y}^{\uparrow}(\mathbf{k})=-v t_a \sin a k_y$, $d_z^{\uparrow} (\mathbf{k})= m + b (\cos a k_x + t^{2}_{a} \cos a k_y)$. Here, $a$ is the lattice constant, $m$ corresponds to the on-site atomic potential, which depends on the local magnetic moment, $v$ describes the hopping amplitude between different orbitals. For simplicity, we set $v=0.1$ eV. The constant $t_a$ introduces a symmetry breaking between the $x$ and $y$ directions within each spin block, rendering the Hamiltonian characteristic of a $d$-wave AM when $t_a \neq 1$. The spin-up and spin-down channels are related by $C_{4z}\mathcal{T}$ or $\mathcal{M}_{xy}$. Here $C_{4z}$ is the fourfold rotation about the $z$ axis, and $\mathcal{M}_{xy}$ is the mirror symmetry. When $t_a = 1$, the system restores $\mathcal{PT}$ symmetry and reduces to  a conventional AFM. The Hamiltonian in Eq.~(\ref{eq1}) describes a QSH insulator when $|b||1-t_a^2| < |m| < |b|(1+t_a^2)$, otherwise, it is topologically trivial. 

For the Hamiltonian in Eq.~(\ref{eq1}), the spin-up and spin-down components are decoupled, their Chern numbers can be evaluated independently\cite{ma2024altermagnetic,gonzalez2025spin,gonzalez2025model}. Here, we denote the Chern numbers of the spin components as $C_{\sigma}$ ($\sigma=\uparrow, \downarrow$), with the total Chern number given by $C = C_{\uparrow} + C_{\downarrow}$, where

\begin{equation}
C_{\sigma} = \frac{1}{4\pi} \int_{\text{BZ}} d\mathbf{k}\frac{\boldsymbol{d}^{\sigma}(\mathbf{k}) \cdot \left[\partial_{k_x} \boldsymbol{d}^{\sigma}(\mathbf{k}) \times \partial_{k_y} \boldsymbol{d}^{\sigma}(\mathbf{k}) \right]}{\left| \boldsymbol{d}^{\sigma}(\mathbf{k}) \right|^3}.
\end{equation}

\begin{figure*}
    \centering
    \includegraphics[width=\linewidth]{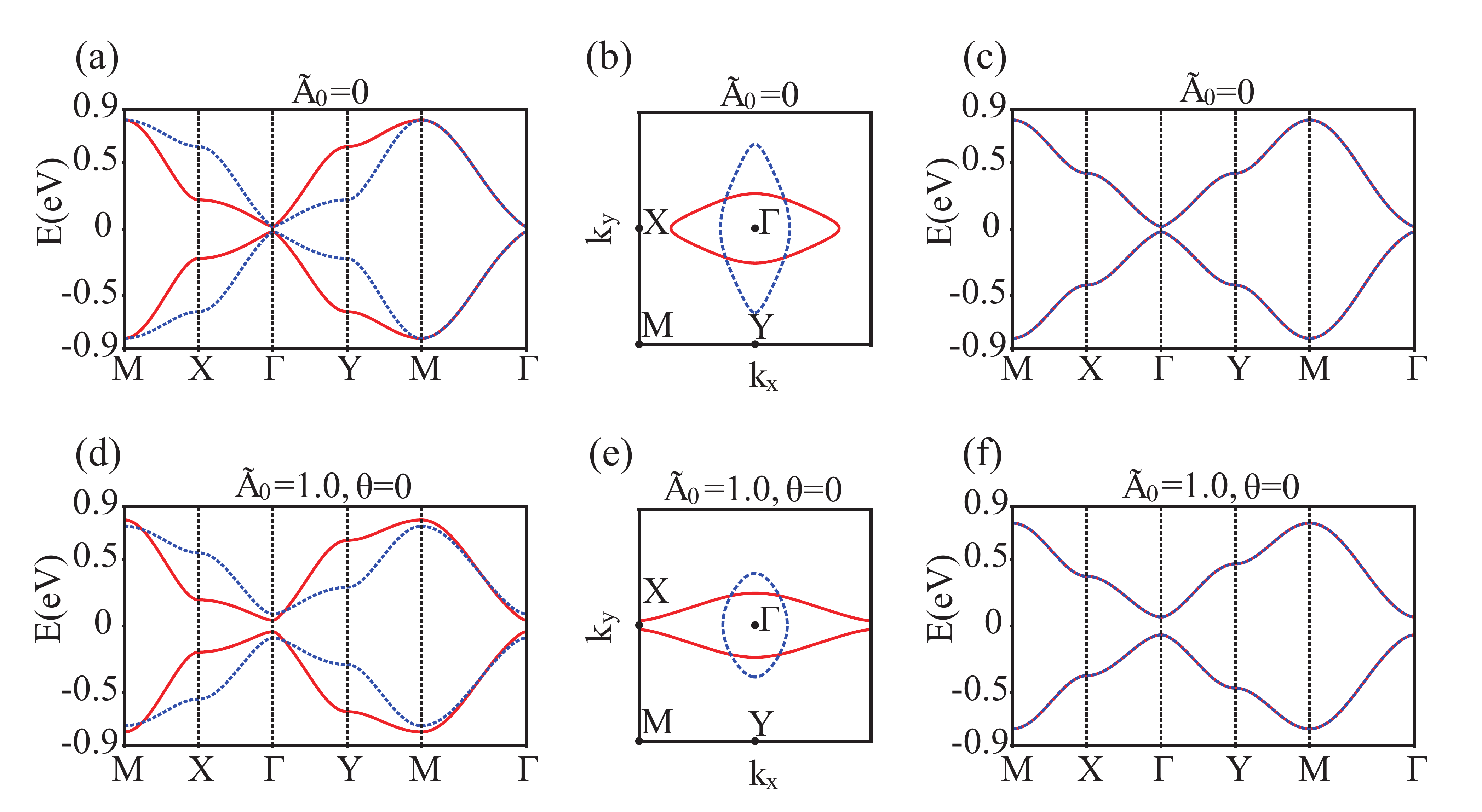}
    \caption{ (a-c) The band structures and the Fermi surfaces in the equilibrium state. (a), (c) The electronic structures for the altermagnet (AM) (a) and the antiferromagnet (AFM) (c). (b) The corresponding  Fermi surfaces at $E_{F}=0.2$ eV for the AM. (d-f) The band structures and the Fermi surfaces under the linearly polarized light irradiation with $\tilde{A_{0}}=1$, $\theta=0$. (d), (f) The electronic structures for the AM (d) and the AFM (f). (e) The corresponding  Fermi surfaces at $E_{F}=0.2$ eV for the AM. Red solid and blue dashed lines denote the spin-up and spin-down bands, respectively. The parameters are $m=0.42$ eV, $b=-0.1$ eV, and $t_a=\sqrt{3}$ for the AM, and $m=0.42$ eV, $b=-0.2$ eV, and $t_a=1$ for the AFM.}
    \label{fig2}
\end{figure*}

\subsection{Effective Hamiltonian with Floquet theory}

For a periodically driven system, the Hamiltonian satisfies $H(\mathbf{k},t)=H(\mathbf{k},t+T)$, where $T=\frac{2\pi}{\omega}$ is the drive period. According to  Floquet theory, the wavefunction of a periodically driven system satisfies $\psi_{\alpha}(\mathbf{k},t)=e^{-i\epsilon_{\mathbf{k}\alpha}t}u_{\alpha}(\mathbf{k},t)$, in which $\epsilon_{\mathbf{k}\alpha}$ is the quasienergy and $u(\mathbf{k},t)$ is periodic, namely, $u_{\alpha}(\mathbf{k},t)=u_{\alpha}(\mathbf{k},t+T)$. The periodic function $u_{\alpha}(\mathbf{k},t)$ satisfies the Schrödinger equation $[H(\mathbf{k},t)-i\partial_t]u_{\alpha}(\mathbf{k},t)=\epsilon_{\mathbf{k}\alpha}u_{\alpha}(\mathbf{k},t)$. Due to the periodicity, $u_{\alpha}(\mathbf{k},t)$ and $H(\mathbf{k},t)$ are  expanded as $u_{\alpha}(\mathbf{k},t)=\sum_{n}u_{n\alpha}(\mathbf{k})e^{in\omega t}$ and $H(\mathbf{k},t)=\sum_{n}H_{n}(\mathbf{k})e^{in\omega t}$, $u_{n\alpha}(\mathbf{k})$ and $H_{n}(\mathbf{k})$ are the $n-$th Fourier component of $u_{\alpha}(\mathbf{k},t)$ and $H(\mathbf{k},t)$. The quasienergy $\epsilon_{\mathbf{k}\alpha}$ satisfies the following equation:
\begin{equation}
\begin{split}
\epsilon_{\mathbf{k}\alpha} u_{m\alpha}(\mathbf{k})=\sum_n [H_{m-n}(\mathbf{k})+m\omega \delta_{mn}]u_{n\alpha}(\mathbf{k}),
\end{split}
\end{equation}
where $H_{m-n}(\mathbf{k})=\frac{1}{T}\int_0^T dtH(\mathbf{k},t)e^{-i(m-n)\omega t}$. In the high-frequency limit, the effective time-independent Hamiltonian can be obtained using high-frequency perturbation theory up to $\mathcal{O}(1/\omega^2)$ \cite{bukov2015universal,kitagawa2011transport,yan2016tunable,mikami2016brillouin,eckardt2015high}.
\begin{equation}
    H_{\text{eff}}(\mathbf{k}) = H_0(\mathbf{k}) + \sum_{n \ge 1} \frac{[H_{n}(\mathbf{k}), H_{-n}(\mathbf{k})]}{n\omega} + \mathcal{O}\left(\frac{1}{\omega^2}\right).
    \label{eq7}
\end{equation}
Now we  consider the irradiation of LPL, with the vector potential given as
\begin{equation}
\bm{A}(t) = A_0 [ \cos{\theta} \cos(\omega t),        \sin{\theta} \cos(\omega t)],
    \label{eq5}
\end{equation}
where  $A_0$ is the amplitude of the LPL, $\theta$ denotes the angle between the polarization direction and the $x$-axis.  The time-dependent Hamiltonian is obtained by making Peierls substitution $\mathbf{k}\rightarrow \mathbf{k}+e\bm{A}(t)/\hbar$. The vector potential of LPL satisfies $\bm{A}(t)=\bm{A}(-t+\tau)$, where $\tau$ is the fractional time translation, which results in the Fourier component $H_n(\mathbf{k}) =e^{-in\omega\tau}H_{-n}(\mathbf{k}) $, the term $\sum_{n \ge 1} \frac{[H_{n}, H_{-n}]}{n\omega}$ vanishes, yielding $H_{\text{eff}}(\mathbf{k}) \approx H_0(\mathbf{k})$. By applying the Jacobi-Anger expansion, the effective model Hamiltonian for the $d$-wave AM  driven by LPL at the high-frequency limit is  derived as:
\begin{equation}
    H_{\text{eff}}(\mathbf{k}) = \begin{pmatrix} 
        H_{\text{eff}}^{\uparrow}(\mathbf{k}) & 0 \\ 
        0 & H_{\text{eff}}^{\downarrow}(\mathbf{k}) 
    \end{pmatrix},
    \label{eq8}
\end{equation}
\begin{equation}
\begin{split}
H_{\text{eff}}^{\sigma}(\mathbf{k}) = \sum _i d^{\sigma}_{i,\text{eff}}(\mathbf{k})\tau_i,
\end{split}
\label{eqhf}
\end{equation}
with  $d_{x,\text{eff}}^{\uparrow}(\mathbf{k}) = -vj_1 \sin a k_x$, $d_{y,\text{eff}}^{\uparrow}(\mathbf{k}) = -v t_a j_2 \sin a k_y$, $d_{z,\text{eff}}^{\uparrow}(\mathbf{k}) = m + b [ j_1 \cos a k_x \  + j_2 t_a^2 \cos a k_y ]$, $d_{x,\text{eff}}^{\downarrow}(\mathbf{k}) = -v j_2 \sin a k_y$, $d_{y,\text{eff}}^{\downarrow}(\mathbf{k}) = -v t_a j_1 \sin a k_x$, $d_{z,\text{eff}}^{\downarrow}(\mathbf{k}) = m + b [ j_2 \cos a k_y  + j_1 t_a^2 \cos a k_x ]$, where  $j_{1}=\mathcal{J}_0(\tilde{A_0}\cos{\theta})$, $j_{2}=\mathcal{J}_0(\tilde{A_0}\sin{\theta})$, $\tilde{A_0}=aeA_{0}/\hbar$ and $\mathcal{J}_0$ is the zeroth-order Bessel function. 

To analyze the minimal $d$-wave altermagnetic model, we restrict our study to the regime of $|m| > \max \left\{ |b|t_a^2, |b| \right\}$ and $m \cdot b<0$. Under this condition, both the conduction band minimum (CBM) and the valence band maximum (VBM) are located in the vicinity of the $\Gamma$ point, and the topological phase transition occurs strictly at the zone center, which provides an ideal platform for capturing Floquet-driven topological responses.

\subsection{Numerical Calculation of the AHC}

In the high-frequency limit, we can calculate the AHC of the effective Hamiltonian (Eq.~(\ref{eq8})). The  AHC is obtained from the following expression:
\cite{nagaosa2010anomalous,xiao2010berry}:
\begin{equation}
\sigma_{xy} = \frac{e^2}{h} \int_{\text{BZ}} \frac{d^2{k}}{2\pi} \sum_{n} f_{n}(\mathbf{k})\Omega_{n}(\mathbf{k}),
\end{equation}
\begin{equation}
\Omega_{n}(\mathbf{k}) = -2 \text{Im} \sum_{m \neq n} \frac{\langle \phi_{n\mathbf{k}} | \hat{v}_{x}| \phi_{m\mathbf{k}} \rangle \langle \phi_{m\mathbf{k}} | \hat{v}_{y} | \phi_{n\mathbf{k}} \rangle}{(\varepsilon_{n\mathbf{k}} - \varepsilon_{m\mathbf{k}})^2},
\end{equation}
where $n$ is band index, $\hat{v}_{x}$=$\frac{1}{\hbar}\frac{\partial H(\mathbf{k})}{\partial k_x} $ and $\hat{v}_{y}$=$\frac{1}{\hbar}\frac{\partial H(\mathbf{k})}{\partial k_y}$ are velocity operators, $f_n(\mathbf{k})$ is the Fermi-Dirac distribution function.  \(\Omega_{n}(\mathbf{k})\) denotes the Berry curvature, \(\varepsilon_{n\mathbf{k}}\) and \(\lvert \phi_{n\mathbf{k}}\rangle\) are the energy eigenvalue and eigenstate of the Floquet effective Hamiltonian, respectively.

\section{Results and Discussion}

\subsection{Modulation of electronic structures by LPL}

We first examine the LPL-induced modulation of the symmetry and electronic structure in AFM and AM. When $t_a = 1$, the system preserves $\mathcal{PT}$ symmetry, representing a conventional AFM.  The system exhibits Kramers degeneracy throughout the entire Brillouin zone (BZ), where the spin-up and spin-down energy bands are completely degenerate[See Fig.~\ref{fig2}(c)]. The spin-up and spin-down bands in the AFM remain degenerate under LPL irradiation due to the preserved $\mathcal{PT}$ symmetry, as shown in Fig.~\ref{fig2}(f). 

When $t_a \neq 1$, the $\mathcal{PT}$ symmetry is naturally broken, and the system transitions into a d-wave AM where the spin-up and spin-down bands are related by the $C_{4z}\mathcal{T}$ or $\mathcal{M}_{xy}$  symmetries. Due to the $C_{4z}\mathcal{T}$ and $\mathcal{M}_{xy}$ symmetries, the spin-up and spin-down bands are  degenerate along the path $\Gamma$-M. However, LPL with arbitrary polarization directions does not preserve the  $C_{4z}\mathcal{T}$ and $\mathcal{M}_{xy}$  symmetries (except when the polarization direction coincides with the mirror plane), lifting the degeneracy along the $\Gamma$-M path, as shown in Fig.~\ref{fig2}(d). Furthermore, the spin-up and spin-down bands are no longer connected by any symmetry operation. This leads to the transition from an AM to a compensated ferrimagnet.

\begin{figure}
    \centering
    \includegraphics[width=\linewidth]{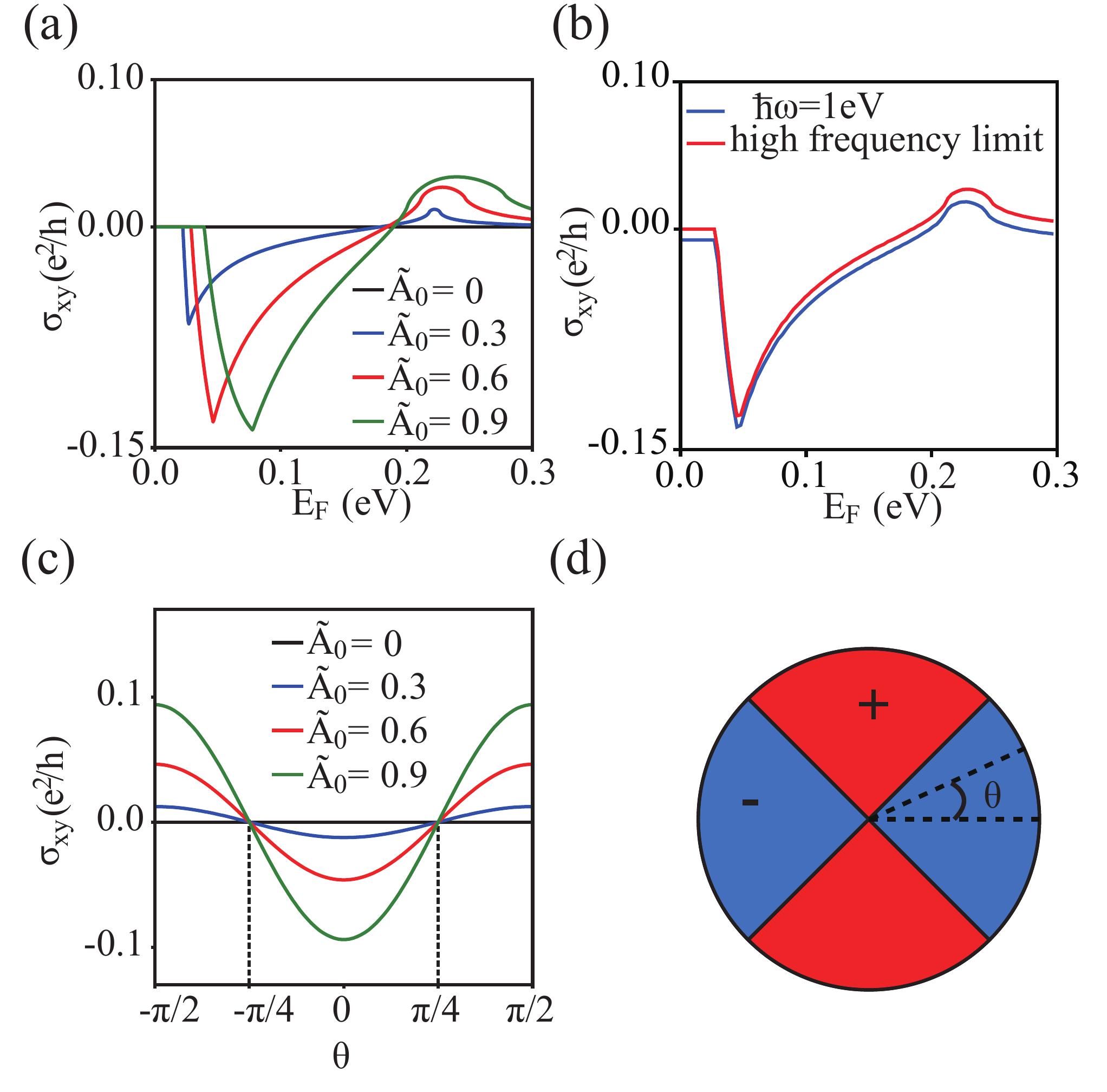}
    \caption{Light-induced anomalous Hall effect (AHE). (a) The anomalous Hall conductivity (AHC) as a function of the Fermi energy for $\tilde{A_{0}}=0$ (black line), $\tilde{A_{0}}=0.3$ (blue line), $\tilde{A_{0}}=0.6$ (red line) and $\tilde{A_{0}}=0.9$ (green line), with $\theta=0$.  (b) The AHC as a function of  $\theta$ at a fixed Fermi energy  $E_{F} = 0.2$ eV. (c) Schematic polar plot of the sign of $\sigma_{xy}$ for  AMs. The  red (blue) regions indicate positive (negative) values of $\sigma_{xy}$. The parameters are $m = 0.42$ eV, $b = -0.1$ eV, and $t_a = \sqrt{3}$.}
    \label{fig3}
\end{figure}

\subsection{LPL-induced AHE}

In the $\mathcal{PT}$-symmetric AFM, $\mathcal{PT}$ symmetry strictly constrains the Berry curvature to zero, so there is no AHE in the $\mathcal{PT}$-symmetric AFM. LPL cannot break $\mathcal{PT}$ symmetry, and therefore LPL cannot induce an AHE in AFM. In the  $d$-wave AM, the spin-up and spin-down electrons are connected by $C_{4z}\mathcal{T}$ or $\mathcal{M}_{xy}$  symmetries enforcing a vanishing AHE. However, LPL can break these  symmetries, thereby inducing an AHE. To evaluate the AHE induced by LPL in AM, two methods were employed to determine the AHC, including a direct computation of the Berry curvature integral within the high-frequency approximation and an alternative approach based on Floquet state occupations. 

In Fig.~\ref{fig3}(a), we calculate the AHC in the AM  as a function of the Fermi level under various LPL intensities within the high-frequency approximation. It is evident that for $\tilde{A_0}=0$, the AHC of the AM vanishes. However, as the light intensity increases, the  $C_{4z}\mathcal{T}$ and $\mathcal{M}_{xy}$  symmetries are broken by the LPL, thereby inducing a finite AHC. The distinct anomalous Hall responses of AFM and AM under LPL can serve as an experimental signature to distinguish between them. 

\begin{figure*}
    \centering
    \includegraphics[width=\linewidth]{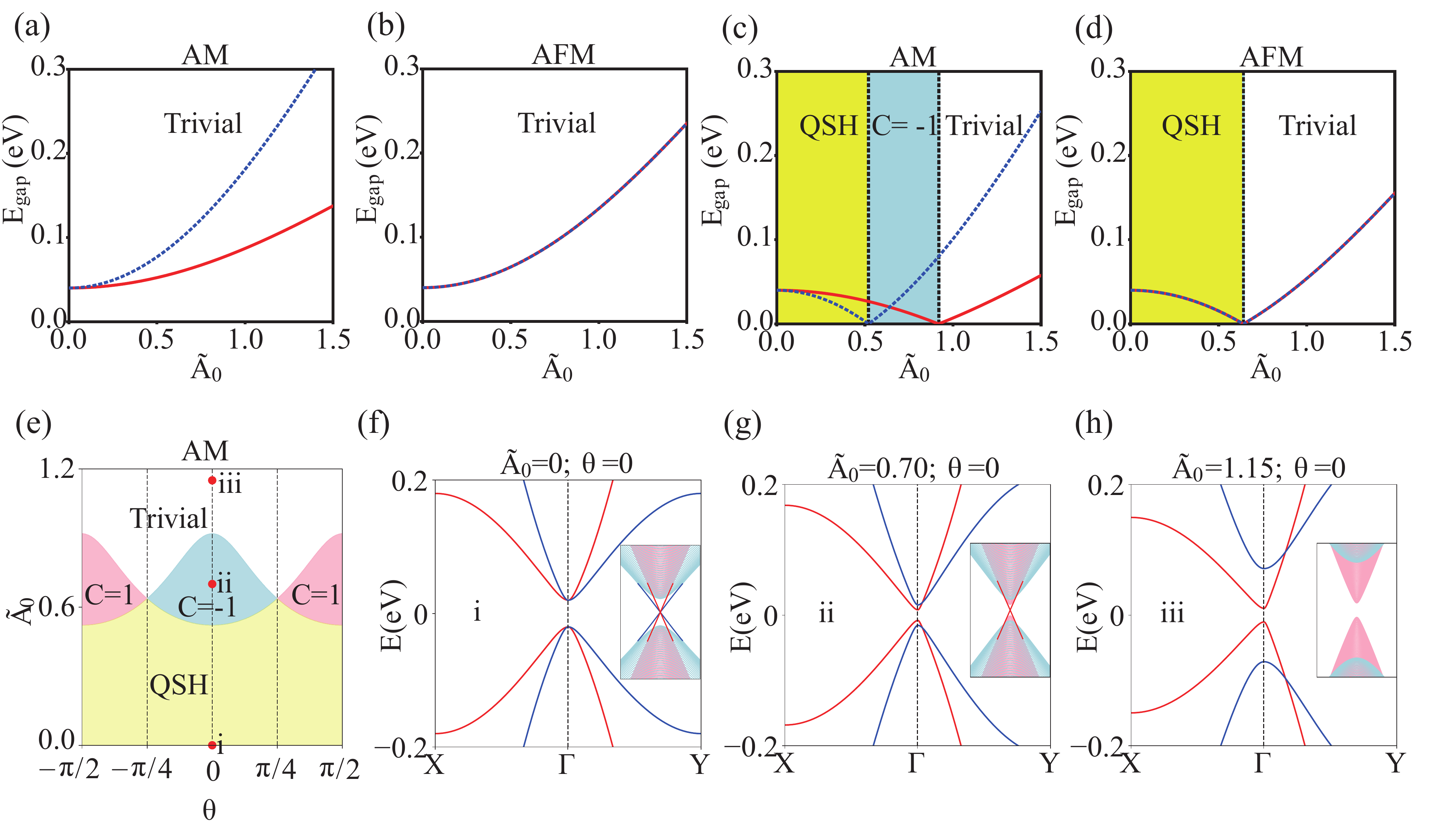}
    \caption{Spin-resolved band gap evolution, topological phase diagram, and band structures under linearly polarized light (LPL) irradiation. (a)-(d) Band gap evolution as a function of the LPL intensity $\tilde{A_0}$ at a fixed polarization direction $\theta=0$ for (a) a trivial altermagnet (AM), (b) a trivial antiferromagnet (AFM), (c) an AM quantum spin Hall (QSH) insulator, and (d) an AFM QSH insulator. Red solid (blue dashed) lines represent the spin-up (spin-down) band gaps. (e) Phase diagram of the $d$-wave AM QSH insulator as a function of $\tilde{A_{0}}$ and $\theta$. Yellow, light blue, pink, and white regions denote the QSH, Chern insulator ($C=-1$), Chern insulator ($C=1$), and topologically trivial phases, respectively. (f)-(h) Bulk band structures of the AM QSH insulator at $\theta=0$ for (f) $\tilde{A_0}=0$, (g) $\tilde{A_0}=0.70$, and (h) $\tilde{A_0}=1.15$, corresponding to the red dots labeled i, ii, and iii in (e), respectively. Red (blue) lines depict the spin-up (spin-down) bands. Insets: Energy bands calculated for a 150-layer ribbon geometry with open boundary conditions along the $x$ direction. Pink (light blue) lines indicate the spin-up (spin-down) bulk bands, while red (blue) lines denote the spin-up (spin-down) edge states. Model parameters are $m=0.42$ eV, $b=-0.1$ eV, and $t_a= \sqrt{3}$ for the trivial AM; $m=0.42$ eV, $b=-0.2$ eV, and $t_a= 1$ for the trivial AFM; $m=0.38$ eV, $b=-0.1$ eV, and $t_a= \sqrt{3}$ for the AM QSH insulator; and $m=0.38$ eV, $b=-0.2$ eV, and $t_a=1$ for the AFM QSH insulator.}
    \label{fig4}
\end{figure*}

In addition to the light intensity, the polarization direction of the LPL serves as another crucial tuning parameter for the AHC. To investigate the influence of the LPL polarization direction on the AHC, in Fig.~\ref{fig3}(b), we calculate the AHC as a function of the polarization angle for various intensities at a fixed Fermi level $E_{F}=0.2$ eV. We observe that the AHE becomes strongly anisotropic with respect to the polarization direction of the LPL, enabling continuous tuning and sign reversal upon polarization rotation.  For a given light intensity $\tilde{A_{0}}$ and Fermi energy $E_{F}$, the induced $|\sigma_{xy}|$ reaches its maximum when the LPL polarization aligns with the $x$- or $y$-axis.  Furthermore, the AHC vanishes at $\theta = \pm \pi/4$, as  $\mathcal{M}_{xy}$  symmetry remains preserved under these conditions. We plot the sign of $\sigma_{xy}$ as a function of $A_0$ and $\theta$ at a fixed $E_{F}$ in a polar coordinate diagram in Fig.~\ref{fig3}(c), where blue represents negative values and red represents positive values. The sign of $\sigma_{xy}$ also exhibits a d-wave characteristic, suggesting that measuring the AHC under LPL irradiation with varying polarization directions offers a means to probe the AM.

\subsection{LPL-induced topological phase transitions}
In this section, we evaluate the topological phase transitions of the $d$-wave AM and AFM under LPL irradiation. The topological phase transitions are usually accompanied by the closing and reopening of the band gap. We first calculate the evolution of the band gaps for spin-up and spin-down bands as a function of LPL intensity for AM and AFM that are initially in trivial or QSH phases, respectively. As shown in Figs.~\ref{fig4}(b) and (d), the gaps for spin-up and spin-down bands remain identical since the AFM retains $\mathcal{PT}$ symmetry under LPL irradiation. In contrast, there is no symmetry linking the spin-up and spin-down channels once the LPL is applied to the AM. As a result, the evolution of the spin-up band gap differs from that of the spin-down band gap as the light intensity changes, as illustrated in Figs.~\ref{fig4}(a) and (c).

If the material is topologically trivial in the absence of light, the energy gaps for both the AM and AFM configurations do not close with increasing light intensity, as shown in Figs.~\ref{fig4}(a) and (b), thus preventing a topological phase transition. For the AFM QSH insulator, as shown in Fig.~\ref{fig4}(d), the band gaps of both spin channels close simultaneously with increasing light intensity, leading to concurrent transitions of $|C_{\uparrow}|$ and $|C_{\downarrow}|$ from 1 to 0. As a result, the QSH phase directly becomes a topologically trivial phase.  Intriguingly, the AM QSH insulator exhibits a different evolution. As illustrated in Fig.~\ref{fig4}(c), the spin-down band gap closes first upon increasing light intensity, driving $C_{\downarrow}$ from 1 to 0 while $C_{\uparrow}$ remains at $-1$. Consequently, the total Chern number becomes $-1$, indicating the emergence of a Chern insulator phase. With further increase of the LPL intensity, the remaining spin-up band gap also closes, eventually driving the material into a trivial phase. Therefore, the AM QSH insulator undergoes a sequential topological transition from the QSH phase to a Chern insulator phase, and finally to a trivial phase. Notably, the nonzero Chern number in this intermediate regime originates entirely from the spin-up bands, identifying it as a spin-polarized Chern insulator. To verify these phase transitions, we calculate the band structures and corresponding edge states for the AM QSH insulator under LPL irradiation. When $\tilde{A_0} = 0$ [Fig.~\ref{fig4}(f)], the system is in the QSH phase, where the spin-up and spin-down channels host  edge states with opposite chiralities. As the light strength is increased to $\tilde{A_0} = 0.7$ [Fig.~\ref{fig4}(g)], the spin-down edge states are destroyed, leaving only the spin-up edge states. The system thus becomes a Chern insulator with $C=-1$, whose edge states are fully spin-polarized.  Upon further increasing the light strength to  $\tilde{A_0} = 1.15$ [Fig.~\ref{fig4}(h)], the edge states disappear, and the system evolves into a topologically trivial insulator.

Fig.~\ref{fig4}(e) presents the topological phase diagram of the AM QSH insulator as a function of the light amplitude $\tilde{A_0}$ and the polarization angle $\theta$. Evidently, the sign of the Chern number in the Chern insulator phase can be switched by varying the polarization angle. Furthermore, when the polarization direction satisfies $\theta=\pm \pi/4$, the AM system follows a transition pathway similar to that of the AFM system, entering the topologically trivial phase directly as the light intensity increases.

\section{Conclusion}

In summary, we employed a tight-binding model that captures two-dimensional out-of-plane AFM  and out-of-plane $d$-wave AM  by tuning model parameters. We employed Floquet theory to explore the optical modulation of their band structures and symmetries by LPL, as well as the emergence of AHE and light-driven topological phase transitions. 

We demonstrate that LPL breaks the $C_{4z}\mathcal{T}$ symmetry connecting the spin-up and spin-down states in AM, while still preserving the $\mathcal{P}\mathcal{T}$ symmetry in AFM. First, within the high-frequency approximation, LPL lifts the spin degeneracy along high-symmetry lines in the effective Hamiltonian of the AM, driving it into a compensated ferrimagnetic state, whereas spin degeneracy remains preserved throughout the entire BZ in the AFM case. Secondly, LPL can induce an AHE in AM but fails in AFM. Thirdly, LPL can drive an AM QSH insulator into a spin-polarized Chern insulator through a topological phase transition, thereby generating a spin-polarized QAH effect. In contrast, LPL cannot induce a nonzero Chern number in a QSH AFM. Furthermore, we observe that the AHE becomes strongly anisotropic with respect to the polarization direction of the LPL, enabling continuous tuning and sign reversal upon polarization rotation. Our findings provide a perspective on the Floquet engineering of AM, as well as a potential method for experimentally distinguishing between AFM and AM materials and probing the spin-splitting pattern of AM.

\section{Acknowledgments}
The work is  supported by the National Key Research and Development  program of China (Grants No. 2024YFA1409100 and No. 2021YFA1400400); the National Natural Science Foundation of Jiangsu Province (Grants No. BK20233001, No. BK20253012 and No. BK20243011), the Natural Science Foundation of China (Grants No. 12534007, No. 92365203, and No. 12504197), and the e-Science Center of Collaborative Innovation Center of Advanced Microstructures, Shandong Provincial Natural Science Foundation (Grant No. ZR2024QA095), and the Fundamental Research Funds for the central Universities (Grant No. 23CX06063A), and the Youth Innovation Team Plan Project for the Higher Education Institution of Shandong Province (Grant No. 2024KJN021).

\bibliography{apssamp}

\end{document}